\documentclass[prd,aps,byrevtex,onecolumn,groupedaddress]{revtex4}%
\usepackage{amsfonts}
\usepackage{graphicx}
\usepackage{amsmath}%
\usepackage{amssymb}
\providecommand{\U}[1]{\protect\rule{.1in}{.1in}}
\begin{document}
\title{Dirac Equation in the Magueijo-Smolin Approach of Double Special
Relativity}
\author{Z. Belhadi$^{1}$, F.\ M\'{e}nas$^{4}$, A. B\'{e}rard$^{2}$, P. Gosselin$^{3}$ and H. Mohrbach$^{2}$}
\affiliation{$^{1}$D\'{e}partement de Physique, Facult\'{e} des Sciences BP 17 Universit\'{e} Mouloud MAMMERI Tizi-Ouzou 15000 Alg%
\'{e}rie}
\affiliation{$^{2}$Institut de Physique, Equipe BioPhyStat, ICPMB, IF CNRS
2843, Universit\'{e} Paul Verlaine-Metz, 1 Bd Arago, 57078 Metz,
Cedex 3, France}
\affiliation{$^{3}$Institut Fourier, UMR 5582 CNRS-UJF, UFR de Math\'{e}%
matiques, Universit\'{e} de Grenoble I, BP74, 38402 Saint Martin d'H%
\`{e}res, Cedex, France}
\affiliation{$^{4}$Laboratoire de Physique et de Chimie Quantique,
Facult\'{e} des Sciences, Universit\'{e} Mouloud Mammeri, BP 17 Tizi Ouzou, Algerie}

\begin{abstract}
We reconsider in details the Dirac equation in the context of the
Magueijo-Smolin approach to the Doubly Special Relativity. Starting from the
deformed dispersion relation we obtain the Dirac equation in momentum space,
allowing us to achieve a more in-depth study of its semiclassical approach.
Finally by means of a deformed correspondence principle we gain access to an
equation in the position space.
\end{abstract}

\maketitle

\section{Introduction}

It is commonly assumed that Heisenberg, intending to remove the electron
self-energy divergence, considered as early as 1930 \cite{PAULI} the
possibility of introducing a space with discrete structure into the
formalism of the quantum mechanics. The symmetry breakdown of this space
however with the non-conservation of the energy-momentum tensor on one hand
and the success of the method based on the renormalization group on the
other, explain why such an approach was abandoned.\ In more recent years the
hypothetical existence of ultra-hight energy cosmic rays particles that
could violate the Greisen-Zatsepin-Kuzmin limit forced physicists to
reconsider the fundamental concepts of the structure of our space-time \cite%
{AMELINO1}. This ideas was also intensely discussed, in particular it was
well known that the Planck length $l_{P}$ plays an important role in quantum
gravity, more specifically in string theory and loops quantum gravity. This
fundamental length is missing in special relativity where two different
observers don't measure the same lengths at the same moment. The fact that $%
l_{P}$ cannot be reached and must be have the same value in all inertial
frames is obviously in contradiction with special relativity, thus
justifying the introduction of a modification of transformations and laws of
this theory. Based on this idea, Amelino-Camelia \cite{AMELINO2} followed by
Magueijo and Smolin \cite{MAGUEIJO1}, pionnered an extended form (Double
Special Relativity\ or DSR) by introducing , in addition to the speed of
light, a second parameter in the form of an energy scale $\kappa =1/$ $l_{P}$
(or Planck energy $E_{p}=\sqrt{\frac{c^{5}\hbar }{G}}=1.956$ $\ 10^{9}$%
Joules $=10^{19}GeV$), that implies a noncommutative space-time structure,
the $\kappa -$ Minkowski space-time. At this level it is important to remark
that, as stated by Amelino Camilia, this short distance scale $l_{P}$ is
relativisticaly fundamental in the same sense as the scale c in Special
Relativity. These two fundamental scales are for example different of $\hbar 
$ which is the scale of the non zero value of the angular momentum, because
they affect the structure of the transformation rules between observers
which is not the case of $\hbar $. In fact these ideas had already been
discussed for several years, though we shall not concern ourselves with the
historical background, which can be found in the recent and very exhaustive
paper of Amelino-Camelia \cite{AMELINO3}. Finally one of the last
generalizations of the DSR is the Deformed General Relativity proposed in 
\cite{GIBBONS} where the geometric framework of an internal De Sitter space
is associated with the non-commutative curved space.

Our purpose in this work is to reconsider in details the Dirac equation in
the context of the Magueijo-Smolin approach to the DSR. In a recent paper 
\cite{NOUS1} we already made a study of this subjet and we found an equation
which is a special case of this one in the context of the Amelino-Camilia
DSR \cite{AGOSTINI}. It is important to state that interesting studies with
this equation have also been conducted, in particular in \cite{HARIKUMAR}
where it was applied to a calculation of the hydrogen atom spectrum was
carried out, or in \cite{DAS} where the discreteness of the generalized
uncertainty principle was used.

We begin in the first part by recalling some consequences of the deformed
Heisenberg algebra, then starting from the deformed dispersion relation we
focus on the obtaining of a Dirac equation in momentum space, allowing us to
achieve a more in-depth study of its semiclassical approach. Finally by
means of a deformed correspondence principle we gain access to an equation
in the position space and by consequently to a deformed current.

\section{Magueijo-Smolin model}

We have choose the Magueijo-Smolin approach mainly because it is one of the
simplest to handle mathematically but also because its results are profound
and far from trivial.

\subsection{Deformed Heisenberg algebra}

To define an energy scale, these workers used a nonlinear action of the
Lorentz group on the momentum space \cite{MAGUEIJO1}, they then determined
how a classical particle in a Minkowski space with a diagonal metric tensor $%
g_{\mu \nu }=(1,-1,-1,-1)$, has its position and momentum which obey the
following deformed Heisenberg algebra defined by means of Poisson brackets

\begin{equation}
\left\{ x^{\mu },x^{\nu }\right\} =\lambda \left( x^{\mu }\delta ^{0\nu
}-x^{\nu }\delta ^{0\mu }\right) ;\left\{ x^{\mu },p^{\nu }\right\} =-g^{\mu
\nu }+\lambda p^{\nu }\delta ^{0\mu };\left\{ p^{\mu },p^{\nu }\right\} =0
\label{Heisengerg algebra}
\end{equation}
where it is usual to introduce the new parameter $\lambda =\kappa ^{-1}.$

Important results are that all the Jacobi identities are conserved and the
invariance of an expression with the quadriposition and the quadri-momentum
under the DSR are given by 
\[
\overline{x}^{2}=g^{\mu \nu }x_{\mu }x_{\nu }\left( 1-\frac{E}{E_{p}}\right)
^{2}=Cte
\]
and 
\[
\overline{p}^{2}=\frac{g^{\mu \nu }p_{\mu }p_{\nu }}{\left( 1-\frac{E}{E_{p}}
\right) ^{2}}=Cte,
\]
but, although the Heisenberg algebra is deformed, the Lorentz algebra is
conserved 
\begin{equation}
\left\{ J^{\mu \nu },J^{\rho \sigma }\right\} =-g^{\mu \rho }J^{\nu \sigma
}+g^{\nu \rho }J^{\mu \sigma }-g^{\nu \sigma }J^{\mu \rho }+g^{\mu \sigma
}J^{\nu \rho }.  \label{Lorentz algebra}
\end{equation}
The position and momentum however are transformed by these Lorentz
transformations under the deformed laws 
\begin{equation}
\left\{ 
\begin{array}{c}
\left\{ x^{\mu },J^{\nu \rho }\right\} =g^{\mu \rho }x^{\nu }-g^{\nu \rho
}x^{\mu }+\lambda \left( p^{\rho }\delta ^{0\nu }-p^{\nu }\delta ^{0\rho
}\right) x^{\mu } \\ 
\left\{ p^{\mu },J^{\nu \rho }\right\} =g^{\mu \rho }p^{\nu }-g^{\nu \rho
}p^{\mu }-\lambda \left( p^{\rho }\delta ^{0\nu }-p^{\nu }\delta ^{0\rho
}\right) p^{\mu }.%
\end{array}
\right.  \label{Lois de transformation}
\end{equation}
We can note that this deformation acting only on the temporal part of the $%
\kappa -$Minkowski space, corresponds only to the action of the Lorentz
boosts and is a particular case of a non-commutative deformation. The
general case acts also on the spacial component by introducing a
quadri-dimensional skew tensor $\theta ^{\mu \nu }$ ($\theta $-Minkowski
space).

\subsection{Magueijo-Smolin Energy dispersion relation}

Starting from the above algebra laws (equations \ref{Heisengerg algebra}, %
\ref{Lorentz algebra} and \ref{Lois de transformation }) they deduced the
following invariant energy dispersion relation \cite{MAGUEIJO1}

\[
\frac{E^{2}-p^{2}c^{2}}{\left( 1-\frac{E}{E_{p}}\right) ^{2}}=\frac{%
E_{0}^{2} }{\left( 1-\frac{E_{0}}{E_{p}}\right) ^{2}}.
\]
Which with the help of the relation 
\[
E_{0}=\frac{mc^{2}}{1+\frac{mc^{2}}{E_{p}}},
\]
becomes 
\begin{equation}
\frac{E^{2}-p^{2}c^{2}}{\left( 1-\frac{E}{E_{p}}\right) ^{2}}=m^{2}c^{4},
\label{Energy MS 1}
\end{equation}
$m$ being called the Casimir mass.

We can observe that this relation is not invariant by the transformation
which reverses the sign of the energy.\ We shall consider this specific
point later. This relation can be transformed in the form 
\begin{equation}
E^{2}=p^{2}c^{2}+m^{2}c^{4}\left( 1-\frac{E}{E_{p}}\right) ^{2},
\label{Energy MS 2}
\end{equation}%
from we obviously find 
\[
E_{\pm }=-\frac{E_{p}M^{2}}{1-M^{2}}\pm \frac{\sqrt{p^{2}c^{2}+\frac{%
E_{p}^{2}M^{2}}{1-M^{2}}}}{\sqrt{1-M^{2}}},
\]%
where the energies of particles and anti-particles are not symmetrical due
to the term $-\frac{E_{p}M^{2}}{1-M^{2}}.$ This remark will be important for
the interpretation of the Dirac equation later on.which is a special form of
the more general relation already studied in many papers \cite{AMELINO3}.
Indeed it should be noted that this equation is a special case of a more
general expression for the dispersion relation that can be written 
\[
E^{2}=p^{2}c^{2}+m^{2}c^{4}+m^{2}c^{4}\left( \frac{E}{E_{p}}\right) ^{n},
\]%
but here the very interesting thing is that the energy dispersion relation
can be separated between energy and momentum. Another important example is
the ultra-high energy case given by 
\[
E^{2}=p^{2}c^{2}+m^{2}c^{4}+p^{2}c^{2}\left( \frac{E}{E_{p}}\right) ^{n},
\]%
with the special case of the toy model of Amelino Camelia \cite{AMELINO3} ($%
n=1$) which is unfortunately not separable 
\[
E^{2}=p^{2}c^{2}+m^{2}c^{4}+p^{2}c^{2}\left( \frac{E}{E_{p}}\right) .
\]%
Returning now to equation (\ref{Energy MS 2}) we very easily find 
\[
\left( 1-M^{2}\right) \left( E+\frac{E_{P}M^{2}}{1-M^{2}}\right)
^{2}=p^{2}c^{2}+m^{2}c^{4}+\frac{E_{P}^{2}M^{4}}{\left( 1-M^{2}\right) },
\]%
where $M=\frac{mc2}{E_{p}}$ so that we can recast this expression in the
usual relation expression of non-deformed special relativity by making the
following change of variables 
\[
\begin{array}{c}
\left\{ 
\begin{array}{c}
\mathcal{M}=\frac{E_{P}M}{c^{2}\sqrt{1-M^{2}}}=\frac{m}{\sqrt{1-M^{2}}} \\ 
\mathcal{E}=\sqrt{1-M^{2}}\left( E+\frac{E_{P}M^{2}}{1-M^{2}}\right) =\left(
E+\frac{\mathcal{M}^{2}c^{4}}{E_{P}}\right) \frac{m}{\mathcal{M}},%
\end{array}%
\right.%
\end{array}%
\]%
so that now 
\begin{equation}
\mathcal{E}^{2}=p^{2}c^{2}+\mathcal{M}^{2}c^{4},
\label{non deformed dispersion}
\end{equation}%
with a new momentum quadrivector $\mathcal{P}=\left( \overrightarrow{p},%
\frac{\mathcal{E}}{c}\right) $ and for solutions 
\[
\left\{ 
\begin{array}{c}
\mathcal{E}_{+}=\sqrt{p^{2}c^{2}+\frac{E_{p}^{2}M^{2}}{1-M^{2}}} \\ 
\mathcal{E}_{-}=-\sqrt{p^{2}c^{2}+\frac{E_{p}^{2}M^{2}}{1-M^{2}}}.%
\end{array}%
\right.
\]%
This result although very simple is not trivial and has an interesting
physical significance as it explains that a relativistic particle in the
Magueijo-Smolin context can be mathematically considered as a particle with
non-deformed energy dispersion relation, but with a deformed energy $%
\mathcal{E}$ and a deformed mass $\mathcal{M}$. Starting from this assertion
we will easily be able to obtain or confirm a certain number of results.

Note that we can easily deduce the non-relativistic limits of the
Magueijo-Smolin energy, already discussed in \cite{CORADDU} 
\[
E_{M-S}=\frac{p^{2}}{2m}+\frac{mc^{2}}{1+\frac{mc^{2}}{E_{p}}}+...,
\]%
and of the new energy 
\[
\mathcal{E}=\frac{p^{2}}{2\mathcal{M}}+\mathcal{M}c^{2}+...=\frac{p^{2}}{2m}+%
\frac{mc^{2}}{\sqrt{1-\frac{m^{2}c^{4}}{E_{P}^{2}}}}+...
\]%
We quickly see that its limits are very different 
\[
\lim_{mc^{2}\rightarrow E_{p}}E_{M-S}\rightarrow \frac{p^{2}}{2m}+\frac{1}{2}%
mc^{2}+...\text{ and }\lim_{mc^{2}\rightarrow E_{p}}\mathcal{E}\rightarrow
\infty .
\]%
In this paragraph our purpose was obviously not to affirm a little as in 
\cite{JAFARI} that, since one can pass via a map from the deformed energy
dispersion relation to a non-deformed energy dispersion relation, DSR is
simply equivalent to the special relativity. We are here in the situation
where the sentence \textquotedblright finding a map between theories
establishes their equivalence\textquotedblright , as Amelino-Camelia wrote
in \cite{AMELINO3}, is not sufficient enough and proves nothing. It is
important to point out that it is only in this \textquotedblright
toy\textquotedblright\ Magueijo-Smolin context that we can easily find a
non-deformed energy dispersion relation, this \textquotedblright kind of
trick\textquotedblright\ is undoubtedly generally impossible. This
opportunity will allow us to go further in our work and to say some new
things about DSR.

\subsection{Non-deformed Lorentz group}

We have seen that the Lorentz group is not deformed, what then happens to
Heisenberg algebra ?

The relation are now given by 
\[
\left\{ x^{\mu },x^{\nu }\right\} =\lambda \left( x^{\mu }\delta ^{0\nu
}-x^{\nu }\delta ^{0\mu }\right) ;\left\{ x^{\mu },\mathcal{P}^{\nu
}\right\} =-G^{\mu \nu }+\lambda \mathcal{P}^{\nu }\delta ^{0\mu };\left\{ 
\mathcal{P}^{\mu },\mathcal{P}^{\nu }\right\} =0,
\]
where we introduced a new mass and Planck energy dependent metric tensor 
\[
G^{\mu \nu }=\left( 
\begin{array}{llll}
\frac{\mathcal{M}}{m} & 0 & 0 & 0 \\ 
0 & -1 & 0 & 0 \\ 
0 & 0 & -1 & 0 \\ 
0 & 0 & 0 & -1%
\end{array}
\right)
\]
which becomes $g^{\mu \nu }$ in the limit of $\mathcal{M}\rightarrow m$ ($%
mc^{2}\ll E_{p}$).

The Jacobi identities are non-deformed and if we put $\mathcal{J}^{\mu \nu
}=x^{\mu }\mathcal{P}^{\nu }-x^{\nu }\mathcal{P}^{\mu }$ , we have the
deformed Lorentz Lie algebra by the ''new metric'' $G^{\mu \rho }$ 
\[
\left\{ \mathcal{J}^{\mu \nu },J^{\rho \sigma }\right\} =-G^{\mu \rho }%
\mathcal{J}^{\nu \sigma }+G^{\nu \rho }\mathcal{J}^{\mu \sigma }-G^{\nu
\sigma }\mathcal{J}^{\mu \rho }+G^{\mu \sigma }\mathcal{J}^{\nu \rho },
\]
and the position and momentum are transformed by these Lorentz
transformations under the deformed laws 
\[
\left\{ 
\begin{array}{c}
\left\{ x^{\mu },\mathcal{J}^{\nu \rho }\right\} =G^{\mu \rho }x^{\nu
}-G^{\nu \rho }x^{\mu }+\lambda \left( \mathcal{P}^{\rho }\delta ^{0\nu }-%
\mathcal{P}^{\nu }\delta ^{0\rho }\right) x^{\mu } \\ 
\\ 
\left\{ \mathcal{P}^{\mu },\mathcal{J}^{\nu \rho }\right\} =G^{\mu \rho }%
\mathcal{P}^{\nu }-G^{\nu \rho }\mathcal{P}^{\mu }-\lambda \left( \mathcal{P}%
^{\rho }\delta ^{0\nu }-\mathcal{P}^{\nu }\delta ^{0\rho }\right) \mathcal{P}%
^{\mu }.%
\end{array}
\right.
\]
We thus obtain the same relations than J.\ Magueijo and L.\ Smolin
(equations \ref{Heisengerg algebra}, \ref{Lorentz algebra} and \ref{Lois de
transformation}), but now with new momentum $\mathcal{P}^{\mu }$ and energy
dependent metric tensor $G^{\mu \nu }.$

\section{Dirac equation in momentum space}

Based on equation (\ref{non deformed dispersion}) one deduces the
non-deformed Dirac equation from a direct standard manner 
\[
\mathcal{E}U(\mathcal{P})=\left( \overrightarrow{\alpha }.\overrightarrow{%
\mathcal{P}}+\mathcal{M}c^{2}\beta \right) U(\mathcal{P}),
\]%
which becomes with the starting variables 
\[
\sqrt{1-M^{2}}\left( E+\frac{E_{p}M^{2}}{1-M^{2}}\right) U\left( p\right)
=\left( c\overrightarrow{\alpha }.\overrightarrow{p}+\frac{E_{p}M}{\sqrt{%
1-M^{2}}}\beta \right) U\left( p\right) ,
\]%
that to say the same equation obtained by a different technique in our last
publication on this subjet \cite{NOUS1} 
\begin{equation}
EU\left( p\right) =\left( \frac{-E_{p}M^{2}}{1-M^{2}}+\frac{c\overrightarrow{%
\alpha }.\overrightarrow{p}}{\sqrt{1-M^{2}}}+\frac{E_{p}M}{1-M^{2}}\beta
\right) U\left( p\right) ,  \label{DDE in momenta}
\end{equation}%
and we thus deduce the deformed Dirac hamiltonian which is hermitian 
\begin{eqnarray}
H_{Dirac}^{Deformed} &=&\frac{-E_{p}M^{2}}{1-M^{2}}+\frac{c\overrightarrow{%
\alpha }.\overrightarrow{p}}{\sqrt{1-M^{2}}}+\frac{E_{p}M}{1-M^{2}}\beta
\label{Dirac hamiltonian} \\
&=&\frac{c\overrightarrow{\alpha }.\overrightarrow{p}}{\sqrt{1-\left( \frac{%
mc^{2}}{E_{p}}\right) ^{2}}}+\frac{mc^{2}}{1-\left( \frac{mc^{2}}{E_{p}}%
\right) ^{2}}\left( \beta -\frac{mc^{2}}{E_{p}}\right) .
\end{eqnarray}%
Interestingly particle anti-particle symmetry breakdown is again reflected
at the Dirac Hamiltonian level by the presence of the term $\frac{-E_{p}M^{2}%
}{1-M^{2}}.$ We have thus retrieve in a very simple manner the Dirac
Hamiltonian found in \cite{NOUS1} by a different procedure. In addition, it
is interesting to note that contrary to the approach developed previously in 
\cite{NOUS1}, the obtained Hamiltonian is directly Hermitian. We shall now
consider some direct applications of this equation.

\subsection{Solving the Dirac equation}

The solution of the non-deformed Dirac equation for a free particle
propagating in the z direction is given in the momentum space by 
\[
U_{a}^{\varepsilon }(p)=\sqrt{\frac{\mathcal{M}c^{2}+\mathcal{E}_{\lambda }}{%
2\mathcal{E}}}\left( 
\begin{array}{l}
\chi _{a} \\ 
\frac{c\sigma _{3}p}{\mathcal{M}c^{2}+\mathcal{E}_{\lambda }}\chi _{a}%
\end{array}%
\right) \exp \left( i\left. \left( (pz-\mathcal{E}_{\lambda }t\right)
\right/ \hbar \right) ,
\]%
where $\lambda =\pm 1$ according to the case of particle or antiparticle and
the spinor field is obviously for $a=1:$ $\chi _{1}=\left( 
\begin{array}{l}
1 \\ 
0%
\end{array}%
\right) $ and for $a=2$ : $\chi _{2}=\left( 
\begin{array}{l}
0 \\ 
1%
\end{array}%
\right) .$

\subsection{Pauli equation}

The consequences of this change in variables is also valid for the Pauli
equation which is the non-relativistic limit of the Dirac equation for an
electron in a magnetic field. Recall that from the standard Dirac equation
we arrive at the relation with an electromagnetics field 
\[
\left( 
\begin{array}{ll}
\pi ^{0}-mc^{2} & -\left( \overrightarrow{\sigma }.\overrightarrow{\pi }%
\right) c \\ 
\left( \overrightarrow{\sigma }.\overrightarrow{\pi }\right) c & -\left( \pi
^{0}+mc^{2}\right)%
\end{array}
\right) \left( 
\begin{array}{l}
\phi (p) \\ 
\chi (p)%
\end{array}
\right) =0,
\]
where 
\[
\pi ^{\mu }=p^{\mu }+eA^{\mu },
\]
particularly

\[
\pi ^{0}=p^{0}+e\frac{V}{c}=m\gamma c+e\frac{V}{c},
\]
which gives in the non-relativistic limit 
\[
\pi ^{0}=mc^{2}+E_{NR}+e\frac{V}{c},
\]
and 
\[
E_{NR}\phi (p)=\left( \frac{\left( \overrightarrow{\sigma }.\overrightarrow{%
\pi }\right) ^{2}}{2m}-eV\right) \phi (p).
\]
In the deformed case we only make the change 
\[
\pi ^{0}=\mathcal{M}c^{2}+\mathcal{E}_{NR}+e\frac{V}{c},
\]
then 
\[
\mathcal{E}_{NR}\phi (p)=\left( \frac{\left( \overrightarrow{\sigma }.%
\overrightarrow{\pi }\right) ^{2}}{2\mathcal{M}}-eV\right) \phi (p),
\]
or 
\[
\mathcal{E}_{NR}\phi (p)=\left( \frac{\left( \overrightarrow{\sigma }.%
\overrightarrow{\pi }\right) ^{2}}{2m}\sqrt{1-\left( \frac{mc^{2}}{E_{p}}%
\right) ^{2}}-eV\right) \phi (p),
\]
which becomes for the hamiltonian 
\[
H_{Pauli}^{Deformed}=\frac{p^{2}}{2\mathcal{M}}+\frac{\hbar }{4\mathcal{M}%
^{2}c^{2}}\overrightarrow{\sigma }\left( \overrightarrow{\nabla }V(R)\wedge 
\overrightarrow{p}\right) +V(R),
\]%
which is now deformed by a $\mathcal{M}$ factor.

\subsection{Deformed Berry phase}

If we consider the adiabatic evolution of the Dirac equation in order to
compute its Berry curvature in momentum space, we found \cite{NOUS2} the
following value of the Berry phase associated with a relativistic particle
submitted to a potential $V(\overrightarrow{r})$ is 
\[
\overrightarrow{a}(\overrightarrow{p},\overrightarrow{\sigma })=\frac{%
c^{2}\hbar \left( \overrightarrow{p}\wedge \overrightarrow{\sigma }\right) }{%
2E\left( E+mc^{2}\right) },
\]%
which in the DRS now becomes 
\[
\overrightarrow{a}_{D}(\overrightarrow{\mathcal{P}},\overrightarrow{\sigma }%
)=\frac{c^{2}\hbar \left( \overrightarrow{\mathcal{P}}\wedge \overrightarrow{%
\sigma }\right) }{2\mathcal{E}\left( \mathcal{E}+\mathcal{M}c^{2}\right) }
\]%
and, if we have made the "standard" change 
\[
\overrightarrow{a}_{D}(\overrightarrow{p},\overrightarrow{\sigma })=\frac{%
\sqrt{1-\lambda ^{2}m^{2}}c^{2}\hbar \left( \overrightarrow{p}\wedge 
\overrightarrow{\sigma }\right) }{2\sqrt{\left( 1-\lambda ^{2}m^{2}\right)
p^{2}c^{2}+m^{2}c^{4}}\left( \sqrt{\left( 1-\lambda ^{2}m^{2}\right)
p^{2}c^{2}+m^{2}c^{4}}+mc^{2}\right) },
\]%
which is the exact result of the publication \cite{NOUS1}. The position
operator is then also very easily to determine 
\[
\overrightarrow{r}_{D}=\overrightarrow{r}+\overrightarrow{a}_{D}.
\]
We now have three kinds of coordinates : $\overrightarrow{r}_{D}$ , $%
\overrightarrow{r}$ and the canonical position $\overrightarrow{R}$

\subsection{Deformed Berry curvature}

For the Berry curvature found in \cite{NOUS2} we have 
\[
\left[ x_{D}^{i},x_{D}^{j}\right] =i\hbar \theta ^{ij}(\overrightarrow{p},%
\overrightarrow{\sigma })=-i\hbar ^{2}\varepsilon ^{ijk}\frac{c^{4}}{2E^{3}}%
\left( m\sigma ^{k}+\frac{p_{k}\left( \overrightarrow{p}.\overrightarrow{%
\sigma }\right) }{E+mc^{2}}\right) ,
\]%
then in DSR 
\begin{eqnarray*}
\overrightarrow{\theta }_{D}(\overrightarrow{\mathcal{P}},\overrightarrow{%
\sigma }) &=&-\hbar \frac{c^{4}}{2\mathcal{E}^{3}}\left( \mathcal{M}%
\overrightarrow{\sigma }+\frac{\overrightarrow{\mathcal{P}}\left( 
\overrightarrow{\mathcal{P}}.\overrightarrow{\sigma }\right) }{\mathcal{E}+%
\mathcal{M}c^{2}}\right)  \\
&=&-\frac{\hbar \left( 1-\lambda ^{2}m^{2}\right) c^{4}}{2\left( \left(
1-\lambda ^{2}m^{2}\right) p^{2}c^{2}+m^{2}c^{4}\right) ^{\frac{3}{2}}}%
\left( mc^{2}\overrightarrow{\sigma }+\frac{\overrightarrow{p}\left( 
\overrightarrow{p}.\overrightarrow{\sigma }\right) }{\sqrt{\left( 1-\lambda
^{2}m^{2}\right) p^{2}c^{2}+m^{2}c^{4}}+mc^{2}}\right) ,
\end{eqnarray*}%
which is also the same result.

\subsection{Deformed dynamic equations with a Berry phase}

From \cite{NOUS2} the dynamic equations are now then written in the
following form 
\[
\left\{ 
\begin{array}{c}
\frac{d\overrightarrow{r}_{D}}{dt}=\frac{\sqrt{1-\lambda ^{2}m^{2}}%
\overrightarrow{p}}{\sqrt{\left( 1-\lambda ^{2}m^{2}\right)
p^{2}c^{2}+m^{2}c^{4}}}-\frac{d\overrightarrow{p}}{dt}\wedge \overrightarrow{%
\theta }_{D} \\ 
\\ 
\frac{d\overrightarrow{p}_{D}}{dt}=\frac{d\overrightarrow{p}}{dt}=-%
\overrightarrow{\nabla }V(\overrightarrow{r}),%
\end{array}
\right.
\]
which gives as direct application the non-relativistic Dirac particle in an
electric potentiel

\[
\left\{ \left. 
\begin{array}{c}
\frac{dX^{i}}{dt}=\frac{p^{i}}{\mathcal{M}}+\frac{e\hbar }{4\mathcal{M}c^{2}}%
\varepsilon ^{ijk}\sigma _{j}\partial _{k}V \\ 
\\ 
\frac{dx^{i}}{dt}=\frac{p^{i}}{\mathcal{M}}+\frac{e\hbar }{2\mathcal{M}c^{2}}%
\varepsilon ^{ijk}\sigma _{j}\partial _{k}V.%
\end{array}%
\right. \right. 
\]

\section{Dirac equation in position space}

We know that position space is fundamental to work in the lagrangian
approach and particularly to study the different symmetries of the physical
problem. But the fact that DSR in momentum space must be extended to
position space is not trivial considering that in this context the Planck
length acts as a fundamental finite length. The interpretation of DSR in
position space then becomes problematic and we do not intend to cover that
topic here. A depper discussion can be found in several papers and for
example a consistent manner is explained in \cite{HOSSENFELDER} or in \cite%
{SMOLIN} where it is shown that this kind of problems occurs only in the
classical picture (with $\hbar \rightarrow 0$) and not in the quantum
picture. Finally we must also announce the two interesting methods for
obtaining the position space of non-linear relativity developed in \cite%
{KIMBERLY}.

It should be recalled that it is only in the Magueijo-Smolin context where
we can find a non deformed energy dispersion relation that we can try to
work in position space so easily. Indeed as we obtained a non-deformed Dirac
equation in momentum space, the temptation was thus strong to switch to
position space with the standard method.

\subsection{Non-deformed Klein-Gordon and Dirac equations}

If we start with the non-deformed energy dispersion relation

\begin{equation}
\mathcal{E}^{2}=p^{2}c^{2}+\mathcal{M}^{2}c^{4},
\end{equation}
which gives by the standard correspondence principle the following
Klein-Gordon equation

\[
\left( -\frac{1}{c^{2}}\frac{\partial ^{2}}{\partial \tau ^{2}}-\left( \frac{%
\mathcal{M}c}{\hbar }\right) ^{2}\right) \phi (\overrightarrow{r},t)=-\Delta
\phi (\overrightarrow{r},t).
\]

\subsubsection{Hermitian hamiltonian}

This last equation implies that the non-deformed Dirac equation in position
space can be written in the fommowing form

\begin{equation}
\left( i\frac{\partial }{\partial \tau }-i\overrightarrow{\alpha }.%
\overrightarrow{\nabla }-\mathcal{M}c^{2}\beta \right) \psi (\overrightarrow{%
r},t)=0,  \label{Non deformed Dirac equation in position}
\end{equation}
where $\tau $ , the time linked to the energy $\mathcal{E}$, is by means of
the definition of $\mathcal{E}$%
\[
\mathcal{E}\rightarrow \frac{\partial }{\partial \tau }=\sqrt{1-M^{2}}\left( 
\frac{\partial }{\partial t}+\frac{E_{P}M^{2}}{1-M^{2}}\right) .
\]
We study now the equation

\[
\left\{ i\left( \sqrt{1-M^{2}}\left( \frac{\partial }{\partial t}+\frac{%
E_{P}M^{2}}{1-M^{2}}\right) \right) -i\overrightarrow{\alpha }.%
\overrightarrow{\nabla }-\mathcal{M}c^{2}\beta \right\} \psi (%
\overrightarrow{r},t)=0,
\]
or

\begin{equation}
i\frac{\partial }{\partial t}\psi (\overrightarrow{r},t)=\left( \frac{%
-E_{p}M^{2}}{1-M^{2}}+\frac{ic\overrightarrow{\alpha }.\overrightarrow{%
\nabla }}{\sqrt{1-M^{2}}}+\frac{E_{p}M}{1-M^{2}}\beta \right) \psi (%
\overrightarrow{r},t),
\label{non deformed transformed Dirac equation in position}
\end{equation}
which gives the hermitian hamiltonian in momentum representation

\[
H_{Dirac}^{Deformed}=\frac{-E_{p}M^{2}}{1-M^{2}}+\frac{c\overrightarrow{%
\alpha }.\overrightarrow{p}}{\sqrt{1-M^{2}}}+\frac{E_{p}M}{1-M^{2}}\beta ,
\]
already find above in the equation (\ref{Dirac hamiltonian}). We found again
retrieve the problem of the difference between particle or antiparticle as
the reverse time symmetry is now broken.

\subsubsection{Continuity equation}

From the non-deformed Dirac equation in positions space (\ref{Non deformed
Dirac equation in position}) we can directly obtain its adjoint equation 
\[
\psi ^{+}(\overrightarrow{r},t)\gamma ^{0}\left( i\hbar \gamma ^{0}\frac{%
\partial }{\partial \tau }-i\hbar \overrightarrow{\gamma }.\overrightarrow{%
\nabla }+\mathcal{M}c^{2}\right) =0,
\]%
where we now use the $\gamma ^{\mu }$ Dirac matrix.\ The continuity equation
is obviously

\[
\frac{\partial \rho (x)}{\partial \tau }+\overrightarrow{\nabla }%
\overrightarrow{j}(\overrightarrow{r},t)=0,
\]
with the usual notations

\[
\left\{ 
\begin{array}{c}
\rho (\overrightarrow{r},t)=\overline{\psi }(\overrightarrow{r},t)\gamma
^{0}\psi (\overrightarrow{r},t) \\ 
\overrightarrow{j}(\overrightarrow{r},t)=\overline{\psi }(\overrightarrow{r}%
,t)\overrightarrow{\gamma }\psi (\overrightarrow{r},t),%
\end{array}
\right.
\]
and the standard notation of Dirac adjoint field

\[
\overline{\psi }(\overrightarrow{r},t))=\psi ^{+}(\overrightarrow{r}%
,t))\gamma ^{0}.
\]%
If we now wish to introduce the \textquotedblright
physics\textquotedblright\ time $t$ we have

\[
\sqrt{1-M^{2}}\left( \frac{\partial }{\partial t}+\frac{E_{P}M^{2}}{1-M^{2}}%
\right) \left( \psi ^{+}(\overrightarrow{r},t)\psi (\overrightarrow{r}%
,t)\right) -\overrightarrow{\nabla }\psi ^{+}(\overrightarrow{r},t)%
\overrightarrow{\alpha }\psi (\overrightarrow{r},t)=0,
\]
then

\[
\frac{\partial }{\partial t}\left( \psi ^{+}(\overrightarrow{r},t)\psi (%
\overrightarrow{r},t)\right) -\overrightarrow{\nabla }\left( \frac{\psi ^{+}(%
\overrightarrow{r},t)\overrightarrow{\alpha }\psi (\overrightarrow{r},t)}{%
\sqrt{1-M^{2}}}\right) =-\frac{E_{P}M^{2}}{\sqrt{1-M^{2}}}\left( \psi (%
\overrightarrow{r},t)\gamma ^{0}\psi (\overrightarrow{r},t)\right) .
\]%
Puting

\[
\left\{ 
\begin{array}{c}
\rho (\overrightarrow{r},t)c=\psi ^{+}(\overrightarrow{r},t)\psi (%
\overrightarrow{r},t)=\left\vert \psi _{0}(\overrightarrow{r},t)\right\vert
^{2}+\left\vert \psi _{1}(\overrightarrow{r},t)\right\vert ^{2}+\left\vert
\psi _{2}(\overrightarrow{r},t)\right\vert ^{2}+\left\vert \psi _{3}(%
\overrightarrow{r},t)\right\vert ^{2} \\ 
\overrightarrow{j}(x)=\frac{\psi ^{+}(x)\overrightarrow{\alpha }\psi (x)}{%
\sqrt{1-M^{2}}},%
\end{array}%
\right.
\]%
we arrive at an equation which tells us that the current is non-conserved 
\[
\frac{\partial }{\partial t}\rho (\overrightarrow{r},t)+div\overrightarrow{j}%
(\overrightarrow{r},t)=-\frac{E_{P}M^{2}}{\sqrt{1-M^{2}}}\rho (%
\overrightarrow{r},t).
\]%
From this it is clear that current is not conserved due to the presence of
the term $-\frac{E_{P}M^{2}}{\sqrt{1-M^{2}}}$ which breaks the particle
anti-particle energy symmetry. Note that, by shifting the energy levels by
the constant term $-\frac{E_{P}M^{2}}{\sqrt{1-M^{2}}},$ we can restate this
symmetry as now $E=\pm \frac{\sqrt{p^{2}c^{2}+\frac{E_{p}^{2}M^{2}}{1-M^{2}}}%
}{\sqrt{1-M^{2}}}$ \ which obviously lead to the conservation of the
current. Nevertheless the 3-D $\overrightarrow{j}(\overrightarrow{r},t)=%
\frac{\psi ^{+}(\overrightarrow{r},t)\overrightarrow{\alpha }\psi (%
\overrightarrow{r},t)}{\sqrt{1-M^{2}}}$ is still deformed. Introducing \ the
velocity as

\[
\overrightarrow{v}=\frac{c}{\sqrt{1-M^{2}}}\overrightarrow{\alpha },
\]
and its eigenvalues by

\[
\overrightarrow{v}=\pm \mathcal{C}\overrightarrow{u},
\]
where $\mathcal{C}=\frac{c}{\sqrt{1-M^{2}}}$ and $\overrightarrow{u}$ is an
unitary vector, the 3-D current as the usual expression $\overrightarrow{j}%
(x)=\rho (x)\overrightarrow{v}.$

We then introduced $\mathcal{C}$ the module of a new kind of
\textquotedblright speed limit\textquotedblright\ which is mass and Planck
energy dependent, the zitterbewegun problem is here also deformed because
the velocity can now be higher than $c$.\ In the limit where $%
mc^{2}\rightarrow E_{p}$, we find that $\mathcal{C}\rightarrow \infty .$ We
could be satisfied with the rescaling of the energy but this would
corresponds to the M-S dispersion relation. Instead we will show that it is
still possible to build another conserved current.

\subsection{Correspondence principle in $\protect\kappa -$Minkowski space}

The question is now to determine whether it is allowed to use a
\textquotedblright correspondence principle\textquotedblright\ in this
context ? The Minkowski phase space obeys the following Heisenberg algebra
with Poisson brackets

\[
\left\{ x^{\mu },x^{\nu }\right\} =0;\left\{ x^{\mu },p^{\nu }\right\}
=-g^{\mu \nu };\left\{ p^{\mu },p^{\nu }\right\} =0.
\]
It is obviously well known that in quantum mechanics we build operators $%
\overset{\wedge }{X}^{\mu }$\ and $\overset{\wedge }{P}^{\mu }$ acting on a
Hilbert space such as

\[
\left\{ 
\begin{array}{c}
\overset{\wedge }{X}^{\mu }\psi (x)=x^{\mu }\psi (x) \\ 
\overset{\wedge }{P}^{\mu }\psi (x)=\frac{\hbar }{i}\frac{\partial \psi (x)}{%
\partial x_{\mu }},%
\end{array}%
\right. 
\]%
and which obey the following Heisenberg algebra with commutators

\[
\left[ \overset{\wedge }{X}^{\mu },\overset{\wedge }{X}^{\nu }\right] =0;%
\left[ \overset{\wedge }{X}^{\mu },\overset{\wedge }{P}^{\nu }\right]
=-i\hbar g^{\mu \nu };\left[ \overset{\wedge }{P}^{\mu },\overset{\wedge }{P}%
^{\nu }\right] =0.
\]%
We have thus achieved the correspondence principle of the quantum mechanics.
In the case of the $\kappa -$ Minkowski space the Heisenberg algebra with
Poisson brackets is

\[
\left\{ x^{\mu },x^{\nu }\right\} =\lambda \left( x^{\mu }g^{0\nu }-x^{\nu
}g^{0\mu }\right) ;\left\{ x^{\mu },p^{\nu }\right\} =-g^{\mu \nu }+\lambda
p^{\nu }g^{0\mu };\left\{ p^{\mu },p^{\nu }\right\} =0.
\]
and then the operators $\overset{\wedge }{X}^{\mu }$\ and $\overset{\wedge }{%
P}^{\mu }$are defined by the generators of the $\kappa -$ Minkowski

\[
\left\{ 
\begin{array}{c}
\overset{\wedge }{X}^{\mu }=x^{\mu }-i\hbar \lambda g^{0\mu }x^{\nu }\frac{%
\partial }{\partial x^{\nu }} \\ 
\overset{\wedge }{P}^{\mu }=\frac{\hbar }{i}\frac{\partial }{\partial x_{\mu
}}.%
\end{array}%
\right. 
\]%
We then arrive after a standard calculus at the following Heisenberg algebra
with commutators

\[
\left[ \overset{\wedge }{X}^{\mu },\overset{\wedge }{X}^{\nu }\right]
=i\hbar \lambda \left( \overset{\wedge }{X}^{\mu }g^{0\nu }-\overset{\wedge }%
{X}^{\nu }g^{0\mu }\right) ;\left[ \overset{\wedge }{X}^{\mu },\overset{%
\wedge }{P}^{\nu }\right] =-i\hbar \left( -g^{\mu \nu }+\lambda \overset{%
\wedge }{P}^{\nu }g^{0\mu }\right) ;\left[ \overset{\wedge }{P}^{\mu },%
\overset{\wedge }{P}^{\nu }\right] =0.
\]%
In conclusion we can see that the standard corespondence principle (case of
the \textquotedblright non deformed theory\textquotedblright ) is associated
with a \textquotedblright deformed correspondence
principle\textquotedblright\ which is true in this context of $\kappa -$%
Minkowski space.

\subsection{Deformed Klein-Gordon and Dirac equations}

If we start with the energy dispersion relation of Magueijo-Smolin

\begin{equation}
E^{2}=p^{2}c^{2}+m^{2}c^{4}\left( 1-\frac{E}{E_{p}}\right) ^{2},
\end{equation}
we can directly deduce, by means of this \textquotedblright deformed
correspondence principle\textquotedblright , the following Klein-Gordon
equation

\[
\left( -\frac{1}{c^{2}}\frac{\partial ^{2}}{\partial t^{2}}-\left( \frac{mc}{%
\hbar }\left( 1-\frac{i\hbar }{E_{p}}\frac{\partial }{\partial t}\right)
\right) ^{2}\right) \widetilde{\phi }(\overrightarrow{r},t)=-\Delta 
\widetilde{\phi }(\overrightarrow{r},t),
\]%
which becomes 
\[
\left( \frac{i}{c}\frac{\partial }{\partial t}-\frac{mc}{\hbar }\left( 1-%
\frac{i\hbar }{E_{p}}\frac{\partial }{\partial t}\right) \right) \left( 
\frac{i}{c}\frac{\partial }{\partial t}+\frac{mc}{\hbar }\left( 1-\frac{%
i\hbar }{E_{p}}\frac{\partial }{\partial t}\right) \right) \widetilde{\phi }(%
\overrightarrow{r},t)=\left( i\overrightarrow{\sigma }.\overrightarrow{%
\nabla }\right) ^{2}\widetilde{\phi }(\overrightarrow{r},t),
\]%
then

\[
\left( \frac{i}{c}\frac{\partial }{\partial t}-\frac{mc}{\hbar }\left( 1-%
\frac{i\hbar }{E_{p}}\frac{\partial }{\partial t}\right) \right) \widetilde{%
\eta }(\overrightarrow{r},t)=\left( i\overrightarrow{\sigma }.%
\overrightarrow{\nabla }\right) \widetilde{\phi }(\overrightarrow{r},t),
\]%
where we have posed

\[
\left( \frac{i}{c}\frac{\partial }{\partial t}+\frac{mc}{\hbar }\left( 1-%
\frac{i\hbar }{E_{p}}\frac{\partial }{\partial t}\right) \right) \widetilde{%
\phi }(\overrightarrow{r},t)=\left( i\overrightarrow{\sigma }.%
\overrightarrow{\nabla }\right) \widetilde{\eta }(\overrightarrow{r},t).
\]%
We can now transform these equations in the following 4D-spinorial equation

\[
\left( 
\begin{array}{ll}
i\frac{\partial }{\partial t} & -i\overrightarrow{\sigma }.\overrightarrow{%
\nabla } \\ 
i\overrightarrow{\sigma }.\overrightarrow{\nabla } & -i\frac{\partial }{%
\partial t}%
\end{array}%
\right) \left( 
\begin{array}{l}
\widetilde{\eta }(\overrightarrow{r},t) \\ 
\widetilde{\phi }(\overrightarrow{r},t)%
\end{array}%
\right) =\frac{mc}{\hbar }\left( 1-\frac{i\hbar }{E_{p}}\frac{\partial }{%
\partial t}\right) ,
\]%
which is written by means of the Dirac matrix $\gamma ^{\mu }$

\begin{equation}
\left( i\gamma ^{0}\frac{\partial }{\partial t}-i.\overrightarrow{\gamma }.%
\overrightarrow{\nabla }-\frac{mc}{\hbar }\left( 1-\frac{i\hbar }{E_{p}}%
\frac{\partial }{\partial t}\right) \right) \widetilde{\psi }(%
\overrightarrow{r},t)=0.  \label{Deformed Dirac equation in position}
\end{equation}%
If we use $\overrightarrow{\alpha }$ and $\beta $ matrix we arrive at the
equation 
\[
\left( \left( 1+\frac{mc}{E_{p}}\beta \right) i\frac{\partial }{\partial t}-i%
\overrightarrow{\alpha }.\overrightarrow{\nabla }+\frac{mc}{\hbar }\beta
\right) \widetilde{\psi }(\overrightarrow{r},t)=0,
\]%
which gives this non-hermitian hamiltonian in momentum representation

\begin{equation}
\widetilde{H}_{Dirac}^{Deformed}=\frac{-E_{p}M^{2}}{1-M^{2}}+\frac{1-M\beta 
}{1-M^{2}}c\overrightarrow{\alpha }.\overrightarrow{p}+\frac{E_{p}M}{1-M^{2}}%
\beta .  \label{non hermitian hamiltonian}
\end{equation}%
which is evidently different of the hermitian one given by the equation (\ref%
{Dirac hamiltonian}). Finally like we easily check that $\left[ \widetilde{H}%
_{Dirac}^{Deformed},J^{i}\right] =0$ , with $J^{i}=L^{i}+\frac{\hbar }{2}%
\Sigma ^{i}$ , this hamiltonian describes well a particle with spin $1/2$.

\subsubsection{Transition from hermitian to non-hermitian hamiltonian}

We have already found in a different context the manner \cite{NOUS1} of
passing from this hermitian to the non-hermitian hamiltonian

\[
\widetilde{H}_{Dirac}^{Deformed}=D^{-1}H_{Dirac}^{Deformed}D,
\]
this hermitic but non-unitary matrix being defined by

\[
\left\{ 
\begin{array}{c}
D=a+b\beta \\ 
D^{-1}=\frac{a-bM}{1-M^{2}}+\frac{b-aM}{1-M^{2}}\beta ,%
\end{array}
\right.
\]
where

\[
\left\{ 
\begin{array}{c}
a=\sqrt{\frac{1+\sqrt{1-M^{2}}}{2}} \\ 
b=\sqrt{\frac{M^{2}}{2\left( 1+\sqrt{1-M^{2}}\right) }}.%
\end{array}
\right.
\]
Naturally we can easily see that the eigenvalues of this two hamiltonians
are real.

\subsubsection{Continuity equation}

From equation (\ref{Deformed Dirac equation in position }) we have now in
the position space the following adjoint Dirac equation

\[
\widetilde{\psi }^{+}(\overrightarrow{r},t)\gamma ^{0}\left( i\left( \gamma
^{0}+M\right) \frac{\partial }{\partial t}-i\overrightarrow{\gamma }.%
\overrightarrow{\nabla }+mc^{2}\right) =0,
\]
we pose then

\[
\overline{\widetilde{\psi }}^{+}(\overrightarrow{r},t=\widetilde{\psi }^{+}(%
\overrightarrow{r},t)\left( \gamma ^{0}+M\right) ,
\]
and by a simple combination of these equations we obtain

\[
\frac{\partial }{\partial t}\left( \overline{\widetilde{\psi }}(%
\overrightarrow{r},t)\gamma ^{0}\widetilde{\psi }(\overrightarrow{r}%
,t)\right) -\overrightarrow{\nabla }.\left( \overline{\widetilde{\psi }}(%
\overrightarrow{r},t)\frac{\gamma ^{0}-M}{1-M^{2}}\gamma ^{0}\overrightarrow{%
\gamma }\widetilde{\psi }(\overrightarrow{r},t)\right) =0,
\]
which is now a conserved current relation

\[
\partial _{\mu }j^{\mu }(\overrightarrow{r},t)=0,
\]
where the charge and current densities are defined by its components

\[
\left\{ 
\begin{array}{c}
\rho (x)=\frac{1}{c}\overline{\widetilde{\psi }}(\overrightarrow{r},t)\gamma
^{0}\widetilde{\psi }(\overrightarrow{r},t)=\frac{1}{c}\widetilde{\psi }^{+}(%
\overrightarrow{r},t)\left( 1+\beta M\right) ^{0}\widetilde{\psi }(%
\overrightarrow{r},t) \\ 
\overrightarrow{j}(x)=\overline{\widetilde{\psi }}(\overrightarrow{r},t)%
\frac{\gamma ^{0}-M}{1-M^{2}}\gamma ^{0}\overrightarrow{\gamma }\widetilde{%
\psi }(\overrightarrow{r},t)=\widetilde{\psi }^{+}(\overrightarrow{r},t)%
\overrightarrow{\gamma }\widetilde{\psi }(\overrightarrow{r},t).%
\end{array}
\right.
\]
As the hamiltonian is not hermitian we have already announced that the
density of charge probability is 
\[
\rho (\overrightarrow{r},t)=\frac{1}{c}\left\{ \left( 1+M\right) \left(
\left| \widetilde{\psi }_{1}(\overrightarrow{r},t)\right| ^{2}+\left| 
\widetilde{\psi }_{2}(\overrightarrow{r},t)\right| ^{2}\right) +\left(
1-M\right) \left( \left| \widetilde{\psi }_{3}(\overrightarrow{r},t)\right|
^{2}+\left| \widetilde{\psi }_{4}(\overrightarrow{r},t)\right| ^{2}\right)
\right\} ,
\]
which is only positive always if: $m\eqslantless \frac{E_{p}}{c^{2}}\backsim
10^{-8}$ Kg.

We have

\[
\overrightarrow{j}(\overrightarrow{r},t)=\rho (\overrightarrow{r},t)%
\overrightarrow{v},
\]
we then deduce the matrix velocity

\[
\overrightarrow{v}=c\frac{\gamma ^{0}-M}{1-M^{2}}\overrightarrow{\gamma }%
=c\left( 
\begin{array}{llll}
\sqrt{\frac{1-M}{1+M}} & 0 & 0 & 0 \\ 
0 & \sqrt{\frac{1-M}{1+M}} & 0 & 0 \\ 
0 & 0 & \sqrt{\frac{1+M}{1-M}} & 0 \\ 
0 & 0 & 0 & \sqrt{\frac{1+M}{1-M}}%
\end{array}
\right) \overrightarrow{\gamma },
\]
which has as the same eigenvalues than in the first case 
\[
v_{i}=\pm \mathcal{C}u_{i}.
\]
The consequences of the zitterbewegun problem are thus the same as in the
preceding case.

To summarize this paragraph, we have seen two manners of attempting to
introduce a continuity equation, one with a hermitian hamiltonian which
leads to a non-conserved current and the other with a non-hermitian
hamiltonian leading to a conserved current.\ These two possibilities result
in the same deformed zitterbewegun problem.

\section{Conclusion}

The main aim of this paper was to demonstrate how in the Magueijo-Smolin DSR
context, we can introduce a non-deformation energy dispersion relation,
deduce a Dirac equation in momentum space and then discuss in the position
space the possibility of a current conservation equation. The other
significant result is the fact that we can work with a non-deformed energy
dispersion relation and the same algebra relations but defined with new
momentum and metric tensor. Finally we easily discovered again the presence
of Berry phase in the adiabatic approximation of the Dirac equation in
momentum space.

The prospects of this approach could be of at least be of two types, firstly
toward a direct study of another equations as Kemmer-Duffin-Petiau or
Feschbach-Villars, secondly toward a generalisation of the type ''Gravity's
rainbow'' of J. Magueijo and L. Smolin \cite{MAGUEIJO2}. To finish we wish
to insist one last time on the simplicity of the deformed energy dispersion
relation which has enabled us to continue the calculus so easily. It is
obviously that the generalisation of this kind of results in the case of a
general non deformed energy dispersion is certainly not trivial.

\textbf{Acknowledgments :} We would like to thank Pr. Subir Ghosh for very
stimulating discussions.

\end{document}